\documentclass[aps,twocolumn,prl,showpacs]{revtex4}
\usepackage{graphicx,amssymb,amsmath}
\usepackage{graphicx}
\usepackage{dcolumn}
\usepackage{bm}

\begin{document}

\title{Breakdown of Particle-Hole Symmetry in the Lowest Landau Level Revealed by Tunneling Spectroscopy}

\author{J.~P. Eisenstein$^1$, L.~N. Pfeiffer$^2$ and K.~W. West$^2$}

\affiliation{$^1$Condensed Matter Physics, California Institute of Technology, Pasadena, CA 91125
\\
$^2$Dept. of Electrical Engineering, Princeton University, Princeton, NJ 08544}

\date{\today}

\begin{abstract} Tunneling measurements on 2D electron gases at high magnetic field reveal a qualitative difference between the two spin sublevels of the lowest Landau level.  While the tunneling current-voltage characteristic at filling factor $\nu = 1/2$ is a single peak shifted from zero bias by a Coulomb pseudogap, the spectrum at $\nu=3/2$ shows a well-resolved double peak structure.  This difference is present regardless of whether $\nu =1/2$ and $\nu = 3/2$ occur at the same or different magnetic fields.  No analogous effect is seen at $\nu = 5/2$ and 7/2 in the first excited Landau level.

\end{abstract}
\pacs{73.43.-f, 73.43.Jn, 71.70.Di} \keywords{tunneling,Landau levels,particle-hole symmetry}
\maketitle
At high magnetic fields and low temperatures electron-electron interactions govern the physics of a two-dimensional electron system (2DES).  If the field $B$ is high enough that only the lowest ($N=0$) orbital Landau level (LL) is occupied by electrons, the fractional quantized Hall effect (FQHE) dominates electrical transport through the system.  Accurately quantized plateaus in the Hall resistance are accompanied by deep minima in the longitudinal resistance at numerous odd-denominator fractional fillings $\nu$ of the Landau level.  If disorder and Landau level mixing effects can be ignored, particle-hole symmetry relates phenomena occuring in the lower spin branch of the $N=0$ LL, i.e. at filling factors $\nu < 1$, to those in the upper spin branch at filling $2 - \nu$ \cite{sym}. Qualitatively, this symmetry appears to hold: Fractional quantized Hall states observed at $\nu =1/3$, 2/3, 2/5, etc. in the lower spin branch are mirrored by robust states at $\nu = 5/3$, 4/3, 8/5, etc. in the upper spin branch. Similarly, compressible composite fermion metallic states appear at both $\nu = 1/2$ and 3/2.  An analogous symmetry is also observed in the excited LLs; the twin FQHE states at $\nu =5/2$ and $7/2$ in the $N=1$ Landau level setting a notable example.

In this paper we examine particle-hole symmetry in two-dimensional electron systems by means of tunneling.  Tunneling is a sensitive probe of many-body effects in electronic systems.  Most famously, the tunneling current-voltage characteristic of superconductor-insulator-superconductor junctions unveils the energy gap due to Cooper pairing in the superconducting electrodes \cite{tinkham}.  In 2DESs, numerous prior studies have demonstrated how tunneling reveals Coulomb correlations \cite{ashoori,jpe,brown,dolgopolov,deviatov}.  In particular, tunneling between the layers in bilayer 2D electron systems clearly shows the importance of both intralayer and interlayer correlations. For closely spaced layers, interlayer tunneling displays spectacular signatures when the total Landau level filling of the bilayer is $\nu_T = 1$ \cite{spielman}.  Reminiscent of the dc Josephson effect, these tunneling features offer compelling evidence for exciton condensation in the bilayer 2D system \cite{jpemacd}.  For widely separated 2DES layers, the case of interest here, the tunneling current-voltage characteristic exposes the convolution of the independent electronic spectral functions of the two layers \cite{theory}. We here make use of this fact to test the validity of particle-hole symmetry in 2D electron systems.  Our results clearly reveal a breakdown of the symmetry between filling factors $\nu = 1/2$ and 3/2 in the lowest Landau level.

The three 2DES samples (A,B, and C) used in this investigation are GaAs/Al$_x$Ga$_{1-x}$As double quantum well (DQW) heterostructures.  Each contains two 20 nm GaAs quantum wells separated by an Al$_x$Ga$_{1-x}$As barrier layer.  The thickness and Al concentration of these barriers are $d_b=20$, 10, and 34 nm, and $x = 0.33$, 1.0, and 0.1, in samples A, B, and C, respectively.  Silicon delta-doping layers positioned above and below the DQW populate the ground subbands of the two quantum wells with nearly identical 2D electron systems. At low temperatures, the nominal density and mobility of these 2DESs are $n = 1.0$, 1.4, and $1.6\times 10^{11}$ cm$^{-2}$ and $\mu = 1.4$, 0.8, and $0.9 \times 10^6$ cm$^2$/Vs, in samples A, B, and C, respectively.  A 250 $\mu$m square mesa, with narrow arms extending outward from its sides to diffused In ohmic contacts, is fabricated on each sample.  A selective depletion scheme is used to allow independent electrical contact to either 2D layer separately \cite{sepcon}.  These independent layer contacts allow for direct measurements of the interlayer tunneling current $I$ and/or tunneling conductance $dI/dV$ resulting from the application of an interlayer voltage $V$.  At zero magnetic field each sample exhibits the sharp tunneling conductance resonances characteristic of momentum and energy conserving tunneling processes \cite{murphy}.  The zero field resonances are also useful in fine-tuning the density balance in the DQW via a large area back gate under the central mesa square of each sample \cite{balance}. (Sample A has an additional large area gate covering the top of the central mesa as well.)  For all the data presented here, the DQW is density balanced at $V = 0$.  Interlayer capacitive effects at $V \neq 0$ disrupt this balance, but the effect is small and, with one possible exception described below, of little consequence here.

\begin{figure}
\includegraphics[width=3.1in, bb=144 106 441 425]{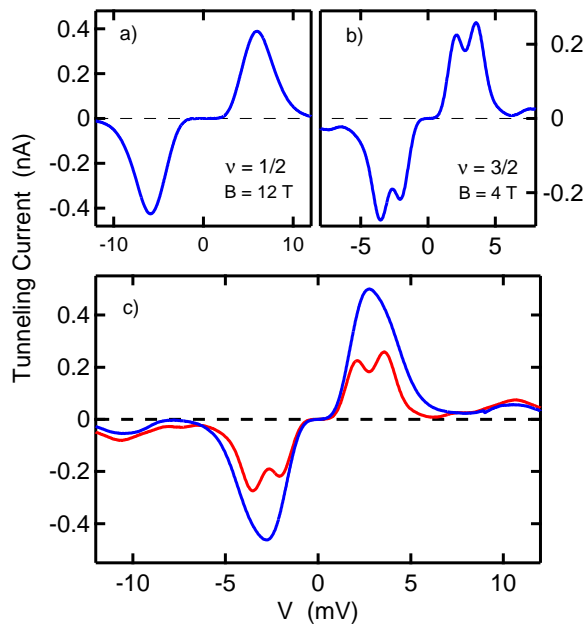}
\caption{\label{}(color online) Tunneling $I-V$ characteristics from sample A. a) $\nu = 1/2$ at $B = 12$ T. b) $\nu = 3/2$ at $B = 4$ T. c) Comparison of $\nu = 3/2$ data at $B =4$ T (same as b)) with $\nu = 1/2$ data at $B = 4.1$ T obtained by gating down the sample density. $T = 0.61$ K for all but $\nu = 1/2$ data in c) where $T = 0.74$ K.}
\end{figure} 
Figure 1(a) displays a tunneling $I-V$ characteristic at from sample A at $B = 12$ T where the Landau level filling factor in each layer is $\nu = hn/eB = 1/2$.  These data exhibit well-known aspects of tunneling into Landau quantized 2D electron systems at high magnetic field.  Most obvious is the strong suppression of tunneling around zero interlayer bias.  This suppression reflects a pseudogap, tied to the Fermi level, in the tunneling density of states. In essence, the pseudogap arises from the inability of the highly correlated electron fluid at high magnetic field to quickly relax the disturbance created by the rapid injection (or extraction) of a tunneling electron. Even if the 2DES is compressible (as it is at $\nu = 1/2$), the non-equilibrium electron states created by tunneling are virtually orthogonal to the ground state of the 2DES with one added (or removed) electron.  As a result, tunneling is heavily suppressed at low energies (and thus low interlayer voltages, $V$). Only when $eV$ is comparable to the mean Coulomb energy between electrons, $e^2/\epsilon \ell$ (with $\ell =(\hbar/eB)^{1/2}$ the magnetic length) at $\nu = 1/2$, does tunneling occur. The width and voltage location of the broad tunneling current peak in Fig. 1(a) have been shown to be consistent with this interaction dominated scenario \cite{exciton}; disorder in the 2DES has little impact in these clean 2D systems. At still higher energies the tunneling current again falls to near zero, only now simply because of the single-particle gap between Landau levels.

Figure 1(b) shows the tunneling $I-V$ characteristic in sample A at $B=4$ T where each 2DES is now at filling factor $\nu = 3/2$.  As at $\nu =1/2$ at $B = 12$ T, a strong suppression of the tunneling around zero interlayer bias is apparent.  At higher voltages, a broad tunneling response is seen, followed again by a collapse in the Landau gap.  In contrast to these similarities between the tunneling spectra in Fig. 1(a) and 1(b) there is one very obvious difference: at $\nu =3/2$ the tunneling response is clearly split into two peaks, while at $\nu =1/2$ only a single peak is seen.  This distinction is seen in all samples discussed here and indeed in all comparable samples that we have examined.

Particle-hole symmetry between $\nu$ and $2-\nu$ applies when the two filling factors are at the same magnetic field. For the data in Fig. 1(a) and 1(b) the density of the 2DES is the same and hance the $\nu = 1/2$ and 3/2 states occur at widely different magnetic fields.  The qualitative difference between the two spectra is therefore not obviously the result of a breakdown of particle-hole symmetry. To address this question directly the densities of the two 2DES layers in sample A were symmetrically reduced, using the gates described above, to produce $\nu = 1/2$ at 4.1 T in the central mesa of the device.  Figure 1(c) compares the tunneling spectrum observed at $\nu = 3/2$ and $B = 4$ T with that observed at $\nu = 1/2$ at essentially the same magnetic field. In spite of the 3-fold reduction in the density, the $\nu = 1/2$ spectrum at $B = 4.1$ T is qualitatively identical to that seen at the same filling factor but at $B = 12$ T. Only a single broad tunneling peak is observed at $\nu = 1/2$ in both cases, in sharp contrast to the split peak found at $\nu = 3/2$.  This fixed field comparison demonstrates a clear breakdown of particle-hole symmetry between these filling factors.

\begin{figure}
\includegraphics[width=3.1in, bb=0 0 239 288]{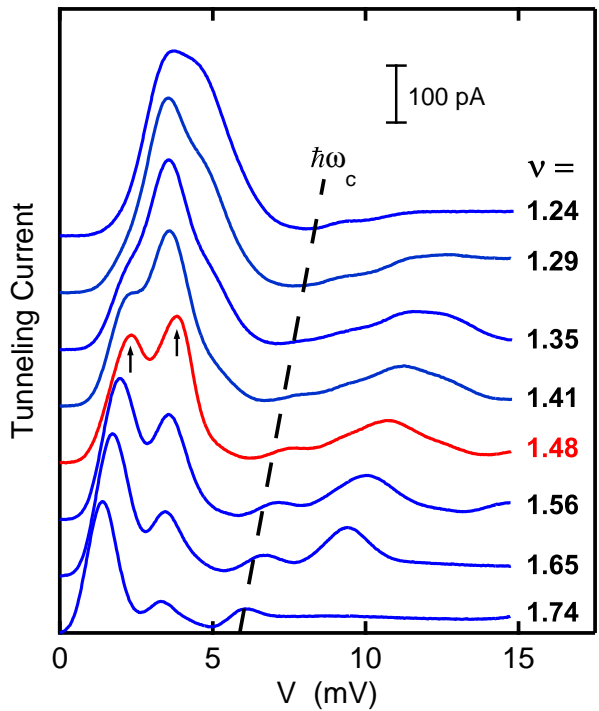}
\caption{\label{}(color online) Tunneling $I-V$ characteristics for various filling factors around $\nu = 3/2$ from sample A at $T = 0.6$ K. Red trace, taken at $B = 4$ T, is closest to $\nu = 3/2$; arrows indicate components of the split peak.  The different data sets are offset vertically for clarity.  Dashed line indicates cyclotron energy $\hbar \omega_c$.}
\end{figure}
The splitting of the tunneling peak observed at $\nu = 3/2$ is not tied precisely to this filling factor.  Figure 2 shows that the splitting is observed over a range of filling factors around $\nu = 3/2$. The relative strength of the two peaks evolves smoothly with $\nu$, with rough equality at $\nu = 3/2$.  Figure 2 also makes clear that additional features are observed at higher energies, beyond the cyclotron energy $\hbar \omega_c$ (indicated by the dashed diagonal line).  This is not surprising, as interaction effects are known to enhance the Landau level splitting beyond the non-interacting cyclotron energy $\hbar \omega_c$ \cite{smith,jpe}.  We note in passing that the width of the main (doublet) tunneling peak just above the zero bias Coulomb gap is roughly $\hbar \omega_c /3$ at the magnetic fields (3.4 $\leq B \leq 4.8$ T) appropriate to Fig. 2.  This too is not surprising since $\hbar \omega_c = e^2/\epsilon \ell$ at 6.3 T in GaAs.

Figure 3 shows two sets of tunneling spectra from sample B for filling factors $\nu < 1$.  The upper set is centered at $\nu = 1/2$ and demonstrates that the single broadened peak in the tunneling $I-V$ curve is insensitive to the precise value of the filling factor.  This is analogous to our observation that the splitting of the $I-V$ curve around $\nu =3/2$ is also insensitive to the precise value of $\nu$.  

Interestingly, as the filling factor deviates further from $\nu = 1/2$, splitting of the tunneling peak does occur.  The lower set of traces in Fig. 3 demonstrates that the splitting, though weak, is clearly evident and is centered roughly around $\nu \approx 0.73$.  Although this observation does not alter our conclusion that particle-hole symmetry between $\nu = 1/2$ and 3/2 is broken, the qualitative similarity between the tunneling spectra at $\nu = 3/2$ and $\nu \approx 0.73$ may offer clues to the origin of the splitting.  This possibility is discussed below.

\begin{figure}
\includegraphics[width=3.1in, bb=0 0 239 288]{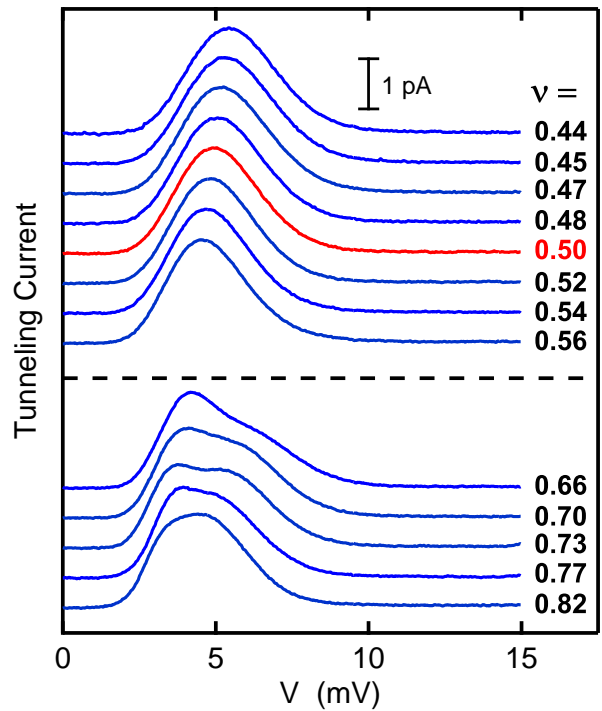}
\caption{\label{}(color online) Tunneling $I-V$ characteristics for various filling factors around $\nu = 1/2$ and $\nu = 3/4$ from sample B at $T = 30$ mK. Red trace is at $\nu = 1/2$ and $B = 11.3$ T.  The different data sets are offset vertically for clarity.}
\end{figure}
The qualitative difference between the tunneling spectra at $\nu = 1/2$ and 3/2 in the $N=0$ LL is not replicated at $\nu = 5/2$ and 7/2 in the $N=1$ first excited LL.  Figure 4 demonstrates this distinction with data from sample C.  In this figure the differential tunneling conductance $dI/dV$ is plotted versus interlayer voltage $V$. Figure 4(a) displays data at $\nu =1/2$ at $B = 13$ T. The strong suppression of the tunneling conductance around $V=0$ is clearly evident.  In place of the single broadened peak seen in the $I-V$ characteristic (e.g. in Fig. 1(a)), the $dI/dV$ spectrum at finite voltage displays a single positive and negative excursion before dropping toward zero.  This is of course expected from differentiating an $I-V$ curve consisting of a single peak.  Figure 4(b) shows the tunneling conductance spectrum at $B = 4.3$ T, very close to $\nu = 3/2$.  Here the spectrum is more complicated; at finite voltage $dI/dV$ oscillates $twice$ before attenuating for $|V| \gtrsim 6$ mV.  This too is expected, provided the $I-V$ characteristic contains two well-resolved peaks.  Hence, the $dI/dV$ data at $\nu = 1/2$ and 3/2 shown in Figs. 4(a) and 4(b) are consistent with the $I-V$ data shown in Figs. 1(a) and 1(b).  

Figures 4(c) and 4(d) show $dI/dV$ data from sample C at $\nu = 5/2$ and 7/2 in the $N=1$ first excited LL, at $B = 2.5$ and 1.83 T, respectively. In both cases the suppression of tunneling around zero bias is again clearly evident although it is weaker than that observed in the $N=0$ LL \cite{suppression}. Moving away from $V=0$ the tunneling conductance at both $\nu = 5/2$ and 7/2 executes a single oscillation before falling toward zero.  This behavior is qualitatively similar to that observed at $\nu = 1/2$ but contrasts sharply with that at $\nu = 3/2$.

\begin{figure}
\includegraphics[width=3.1in, bb=161 95 455 344]{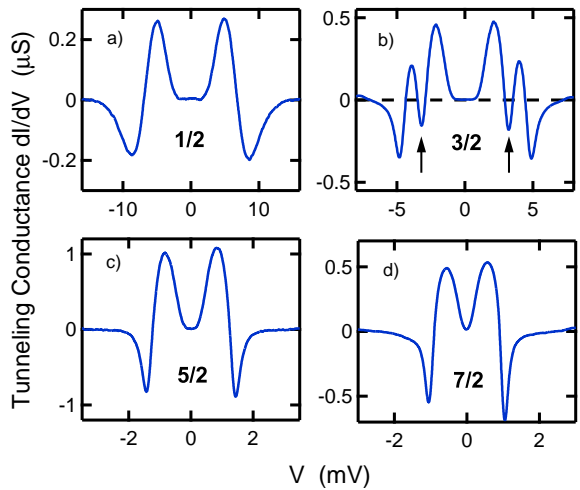}
\caption{\label{}(color online) Tunneling conductance data at $\nu = 1/2$, 3/2, 5/2, and 7/2 from sample C.  Only at $\nu = 3/2$ is an extra oscillation (indicated by arrows) observed.}
\end{figure}
One possible mechanism for the splitting of the tunneling $I-V$ characteristic at $\nu = 3/2$ involves the spin polarization of the 2DES.  If, for the moment, we assume that at $\nu = 3/2$ the lower spin branch of the $N=0$ LL is fully filled while the upper branch is one-half filled, then, provided that individual tunneling events are spin preserving, the tunneling $I-V$ curve would reflect only the spectral function of the upper spin branch.  A zero bias Coulomb gap would be expected but there would be no obvious reason for the peak in the tunnel current at finite bias to be split.  If, however, $both$ spin branches of the $N=0$ LL are partially filled at $\nu = 3/2$, then electrons of both spin directions could tunnel.  The spectral functions of each spin population would possess Coulomb gaps at the Fermi level, but the magnitude of these gaps and the width of the spectral distributions about them would in general be unequal.  In this case spin-preserving tunneling would be suppressed at zero bias as usual, but a split tunneling peak at finite bias could easily result from the different spectral distributions of the two spins.  

There is by now copious evidence that the spin polarization of the 2DES in typical GaAs heterostructures does not follow simple Pauli counting rules in the lowest Landau level.  For example, it is well known that the FQHE at $\nu = 2/3$ can exist in both spin polarized and unpolarized states \cite{twothirds}.  If the splitting of the tunneling $I-V$ curve we observe around $\nu = 3/2$ is due to spin effects, then the weaker splitting shown in Fig. 3b around $\nu \approx 0.73$ might in turn reflect the spin state of the nearby $\nu = 2/3$ FQHE \cite{capac}. In fact, there is strong evidence for a lack of maximal polarization at both $\nu = 3/2$ in the upper spin branch and $\nu = 1/2$ in the lowest LL at appropriately low magnetic fields \cite{stormer,tracy}.  While this would seem to assist in understanding the split tunneling peak seen at $\nu =3/2$, it does not solve the mystery of the broken particle-hole symmetry so clearly displayed in Fig. 1(b).  If particle-hole symmetry is present the net spin magnetization at $\nu = 3/2$ and $\nu = 1/2$ must be the same (when compared at the same magnetic field).  If the splitting of the tunnel peak at $\nu = 3/2$ is due to less than maximal spin polarization, the same splitting should be seen at $\nu = 1/2$.  Figure 1(b) shows that this is clearly not the case.  

Both random disorder and mixing between Landau levels can break particle-hole symmetry.  Since the splitting of the tunneling $I-V$ curve at $\nu = 3/2$ is observed in all of the present samples (plus all of the very many comparable samples we have ever examined) extrinsic disorder, which varies significantly from one sample to the next, would seem to be an unlikely source of it.  In contrast, Landau level mixing is an intrinsic effect.  As already mentioned, the mean Coulomb energy $e^2/\epsilon\ell$ equals the Landau level splitting $\hbar \omega_c$ at $\sim 6$ T in GaAs, so it is likely that Landau level mixing is important at $\nu = 3/2$ in typical samples.  At fixed magnetic field the $\nu = 3/2$ state in the upper spin branch of the $N = 0$ LL is closer in energy to the unoccupied $N = 1$ LL than is the $\nu = 1/2$ state in the lower spin branch, thus raising the prospect for stronger mixing at $\nu = 3/2$ than at $\nu = 1/2$.  Whether such mixing effects can produce the split tunneling spectra observed at $\nu = 3/2$ is so far unknown.

In conclusion, we have used tunneling spectroscopy to uncover a clear violation of particle-hole symmetry in GaAs-based two-dimensional electron systems at high magnetic field.  The tunneling $I-V$ characteristics at $\nu = 3/2$ in the lowest Landau level show a clear doublet structure which is not seen at the conjugate $\nu = 1/2$ state.  No similar breaking of particle-hole symmetry has been found between $\nu = 5/2$ and 7/2 in the second Landau level.

This work was supported by Bell Laboratories.  We thank Allan MacDonald for useful discussions.


\begin{references}
\bibitem{balance} Density balance is readily detected via the symmetry of the tunneling resonance at $B = 0$.
\bibitem{suppression} That the zero bias suppression is weaker at $\nu = 5/2$ and 7/2 than at $\nu =1/2$ and 3/2 is unsurprising; at $\nu = 5/2$ and 7/2 the density of interacting electrons in the partially filled valence Landau level is only 1/5 and 1/7, respectively, of the total.
\bibitem{capac} Indeed, owing to the capacitance between the layers, at $V \approx 5$ mV, where the splitting occurs, the filling factors of the individual layers in sample B are about $\nu_1 = 0.67$ and $\nu_2 = 0.79$, thus enhancing the prospect that the $\nu_1 = 2/3$ FQHE is involved.
\end{references}
\end{document}